\documentclass[%
 aip,cha,
 amsmath,amssymb,
preprint,%
floatfix,
author-numerical,%
]{revtex4-1}

\usepackage{graphicx}% Include figure files
\usepackage{dcolumn}% Align table columns on decimal point
\usepackage{bm}% bold math
\begin{document}

\preprint{AIP/123-QED}

\title{Long-range fluctuations and multifractality in connectivity density time series of a wind speed monitoring network}

\author{Mohamed Laib}
 \email{Mohamed.Laib@unil.ch}
 \affiliation{IDYST, Faculty of Geosciences and Environment, University of Lausanne 1015, Switzerland.} %Lines break automatically or can be forced with \\
\author{Luciano Telesca}%
 \affiliation{CNR, Istituto di Metodologie per l'Analisi Ambientale, 85050 Tito (PZ), Italy.}

\author{Mikhail Kanevski}
\affiliation{%
IDYST, Faculty of Geosciences and Environment, University of Lausanne, Switzerland.%\\This line break forced% with \\
}%

\date{\today}% It is always \today, today,

\begin{abstract}
This paper studies the daily connectivity time series of a wind speed-monitoring network using multifractal detrended fluctuation analysis. It investigates the long-range fluctuation and multifractality in the residuals of the connectivity time series. Our findings reveal that the daily connectivity of the correlation-based network is persistent for any correlation threshold. Further, the multifractality degree is higher for larger absolute values of the correlation threshold.

\end{abstract}

%Use showkeys class option if keyword
\maketitle

\begin{quotation}
Recent breakthroughs in the technology of collecting and storing data  have enhanced meteorological monitoring systems, enabling them to record meteo-climatic parameters with greater frequency. These monitoring systems are currently constituted by a multitude of sensors scattered in space. 
As consequences, studying the interaction between sensors is becoming very challenging. Especially, if the monitoring system is characterized by the presence of diverse topographies like the Swiss Monitoring system.
This work proposes a correlation-based network to study interaction among the sensors. The daily evolution of the connectivity density of this network is analysed using a multifractal detrended fluctuation analysis. The results of this study could open new methodological avenues for studies devoted to the analysis of wind speed time series.

\end{quotation}

\section{Introduction}
\label{intro}

Due to continuous improvements in the technology of measuring, collecting and storing data, environmental phenomena are now being recorded with higher frequency than in the past. In many cases, the phenomena being measured are multivariate, complex, and non-linear. As a consequence, they need to be analysed using robust and advanced statistical methods to be deeply understood. If a meteo-climatic phenomenon is monitored over time using multiple sensors distributed in space;  the temporal evolution of the interaction among all the units becomes more complex. Therefore, it is important to investigate how all measured times series correlate with each other across time.

In recent decades, complex networks have been gaining popularity as a way of understanding meteo-climatic parameters measured in different places in space. In fact, several studies have been conducted using network analysis\cite{moreno}. For instance, in investigations of atmospheric systems \cite{Tsonis2006},studying the characteristics of El-Ni\~no and La-Ni\~na phenomena  \cite{Tsonis2008}, modelling and characterizing climate systems \cite{Donges2009a, Donges2009b, Gozolchiani2008}. There are different approaches in constructing a network, the most commonly used is to grid a given field (temperature, wind, air pressure $\ldots$) and take the grid cells as the nodes of the network where the edges are defined by certain relationships. Such relationships can be based on linear or non-linear dependency metrics \cite{Tsonis2004, Yamasaki2008, Tsonis2008, Donges2009b, Steinhaeuser2009}.

This paper investigates a wind speed-monitoring system in Switzerland  using complex networks. Wind phenomenon, as an important feature of climate system, has been studied using several statistical tools and techniques, like extreme value theory and copula \cite{DAMICO2015}, machine learning algorithms \cite{Treiber2016}, visibility graphs \cite{PIERINI2012}, Markov chains \cite{KANTZ2004}, fractal analysis \cite{DEOLIVEIRASANTOS2012, Fortuna2014}, multifractal analysis \cite{Telesca2016, Garcia2013}. In this work, the used network approach considers each wind station as a node of the network, and the edges linking two different nodes are weighted by the Pearson correlation coefficients between them. An edge is only considered to exist if the corresponding correlation coefficient is above/below a fixed threshold. Therefore, the topology of the network is given by the set of nodes that are interconnected by the existing edges.

This study focuses on the investigation of the temporal fluctuations of the  connectivity density (defined in the next section), which is employed to describe the network topology. The connectivity density is capable of depicting the degree of nodes interconnected on the base of the Pearson correlation. This paper focuses on a daily time-varying network, and studies the connectivity density changing each day (due to the daily changing correlation between any two nodes of the network). Thus, a daily time series of the connectivity density is obtained. Like any natural time series, the connectivity time series of the wind speed monitoring system is characterized by temporal fluctuations. The analysis performed using robust statistical methods is able to reveal important dynamical features that enabled us to get better insight into the time dynamics of the wind speed. Many natural phenomena have been widely analysed by using fractal and multifractal methods, due to their capability to highlight long-range correlation properties, scaling characteristics, persistence, etc. In this study, the multifractal detrended fluctuation analysis (MFDFA) is applied to detect the presence of long-range properties in the connectivity density time series, and to quantify possible intermittent signatures linked with heterogeneous dynamics. The paper combines three different aspects of the same phenomenon  (such as wind speed):
\begin{itemize}
\item spatial: due to the distribution of 119 sensors all over Switzerland. And the topographic characterization of the country;
\item temporal: due to   the analysis of the network topology on a daily basis;
\item correlative: due to the network topology, which is defined based on the correlation threshold.
\end{itemize}

Other researchers have employed fractals in analysing the inner time structure of wind speed. For instance, Chang et al. \cite{CHANG2012} analysed the box-counting fractal dimension of wind speed recorded at three wind farms in Taiwan with different climatic conditions. They found an inverse correlation between the mean wind speed and the fractal dimension. Long-term correlation properties of wind speed records were quantified by De Oliveira Santos et al.  \cite{DEOLIVEIRASANTOS2012}, who applied the detrended fluctuation analysis (DFA) to average and maximum hourly wind speed time series measured at four weather stations in Brazil. Their results revealed different scaling regimes. Fortuna et al.  \cite{Fortuna2014} calculated the Hurst exponent $H$ to quantify the persistence in wind speed. Kavasseri and Nagarajan \cite{Kavasseri2005} investigated hourly means of wind speed in USA, finding that the binomial cascade multiplicative model fit the data well. Telesca and Lovallo \cite{telesca2011} analysed the hourly wind speed time series at several heights from the ground, with a range between 50m and 213m. They found that most of the multifractality of the wind speed was due to the different long-range correlations for small and large speed fluctuations. De Figueiredo et al.  \cite{DEFIGUEIRA2014} applied the MFDFA to the mean and maximum of four wind speed time series in Brazil and found that both persistently correlated with a larger multifractality for the maximum than for the mean. Piacquadio and de la Barra used some multifractal parameters of wind speed as local indicators of climate change  \cite{PIACQUADIO2014}. In these previous studies, the fractal/multifractal concept was applied directly to wind speed time series. However, to our knowledge, no research has applied the fractals/multifractals to the time variation of a parameter that explains the network topology of a wind speed monitoring system.

This study contributes to enriching understandings of wind processes in complex environments like Switzerland. In fact, knowledge of the dynamic interactions among the different stations of a spatially wide wind monitoring system, and comprehension of the long-range and intermittent features of these interactions, would help in the planning and design of monitoring systems.
%%%%%%%%%%%

\section{Data and network construction}
\label{sec:1}

\begin{figure}[h]
%%\rule{1cm}{1cm}width=\linewidth
\centering
\includegraphics[width=\linewidth]{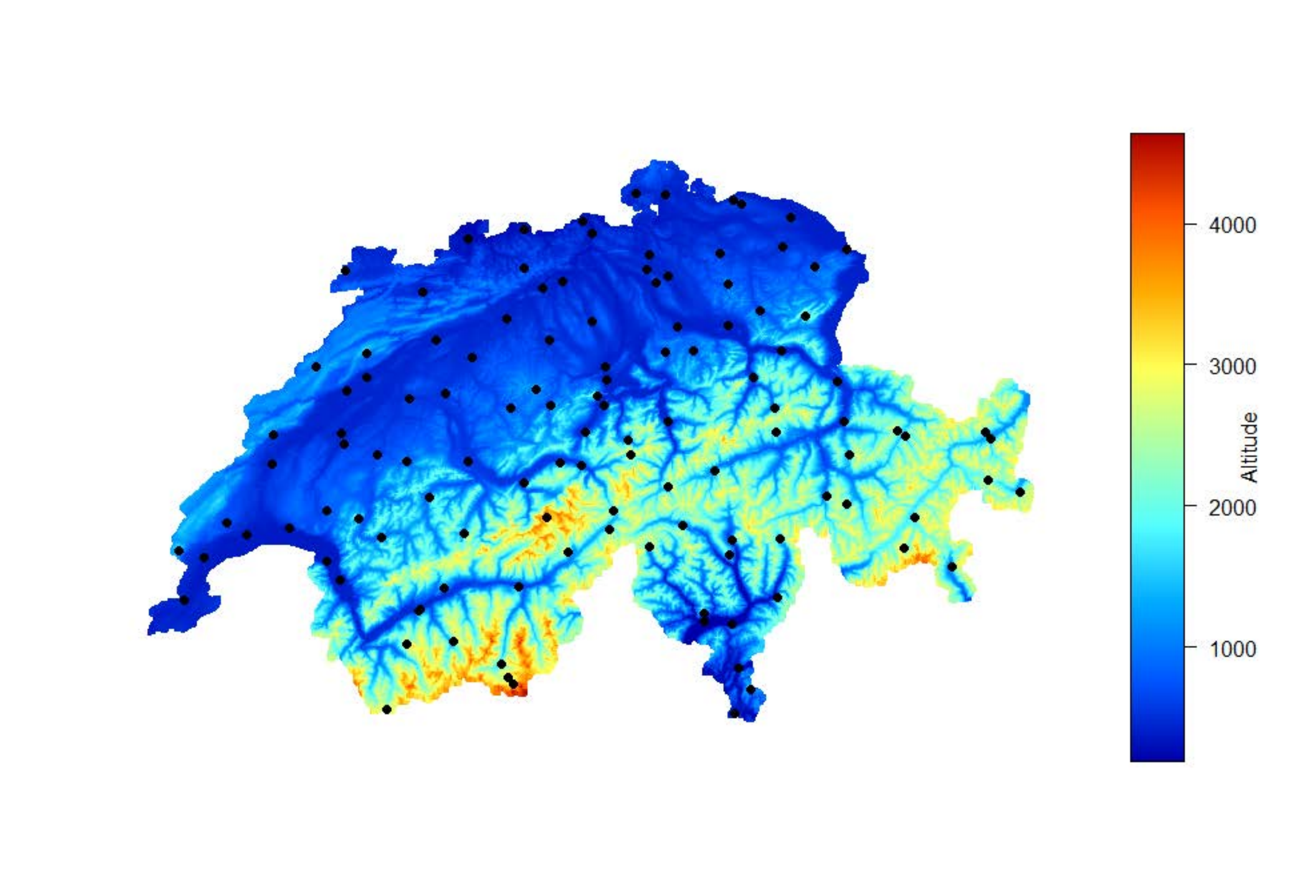}
\caption{\footnotesize Study area and location of wind measurement stations.}
\label{fig1}  
\end{figure}

\begin{figure}
%%\rule{1cm}{1cm}width=\linewidth
\includegraphics[width=\linewidth]{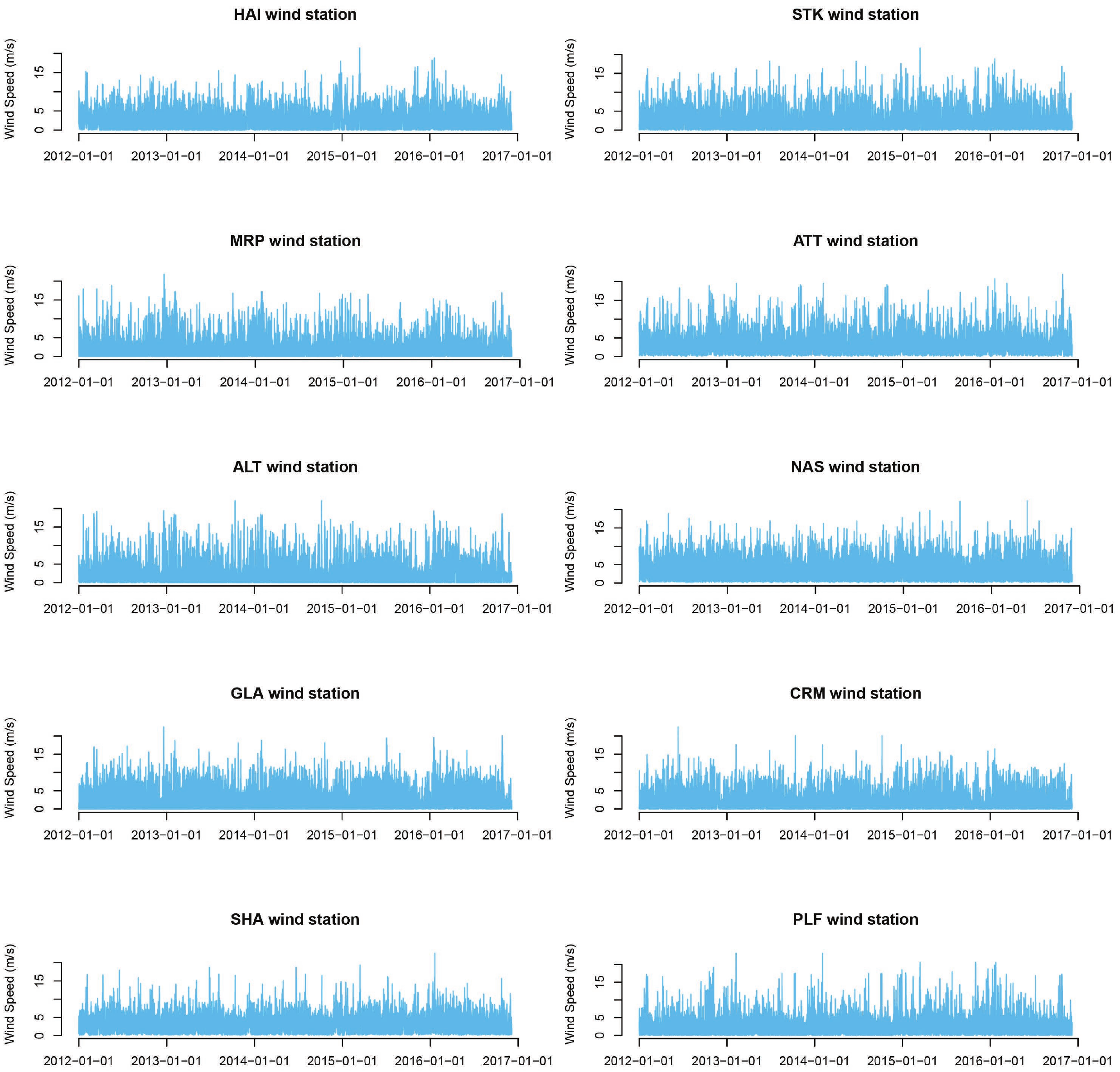}
\caption{\footnotesize Some wind speed time series.}
\label{fig2}  
\end{figure}

The Federal Office of Meteorology and Climatology of Switzerland (MeteoSwiss) manages a dense monitoring system of different meteorological parameters. This system is distributed over the entire country (Fig. \ref{fig1}). The data used in this paper consist of high frequency wind speeds (10-min sampling), recorded in 119 places between 2012 and 2016. Fig. \ref{fig2} shows some of wind speed series.

\begin{figure}
%%\rule{1cm}{1cm}width=\linewidth
\includegraphics[width=\linewidth]{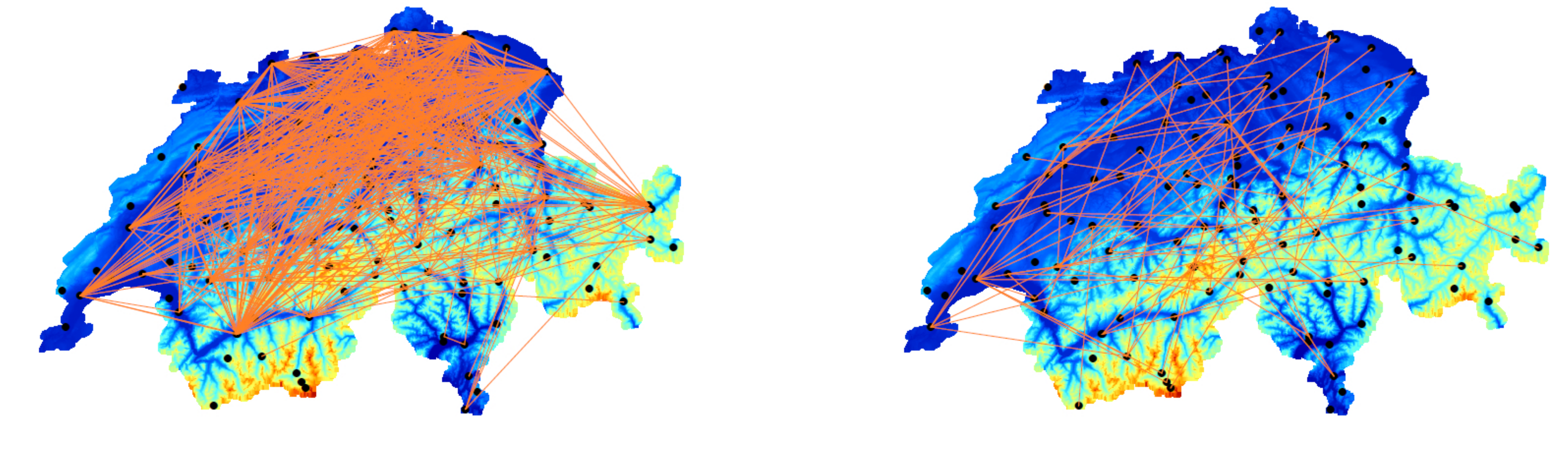}
\caption{\footnotesize Networks on two different days considering a threshold of 0.7: April 9, 2012 with $\Delta=0.088$ (left) and September 4, 2016 with $\Delta=0.012$ (right).}
\label{fig3}  
\end{figure}

As mentioned above, we constructed a daily varying correlation network for wind speed. The wind measuring stations as nodes and the edges are defined by the Pearson correlation:

\begin{equation}
\rho_{XY}=\frac{\sum(x_{i}-\bar{x})\: \sum(y_{i}-\bar{y})}{\sqrt{{\sum(x_{i}-\bar{x})}^2} \: \sqrt{{\sum(y_{i}-\bar{y})}^2}}
\end{equation}
where $\bar{x}$, $\bar{y}$ denote the mean of two wind time series $X$,$Y$ respectively. The coefficient of correlation, as is well known, quantifies the existing relationships between two random variables. The network topology depends on the choice of a threshold; an edge only exists if $\rho$ is below/above the fixed threshold. This topology varies day by day because each day the amount of the existing edge changes due to the daily changing correlation between any two nodes.  Fig. \ref{fig3} shows two different network topologies on two different days based on a threshold of 0.7, and all the edges whose weight is above this threshold.

In this paper, we not only consider positive values of the thresholds but also negative values. When the threshold is positive, the network considers all the edges whose weight is above the threshold, while when the threshold is negative, the network considers all the edges whose weight is below the threshold. For zero threshold, we built two networks, one considering all the edges whose weight is above zero, and the other considering all the edges with a weight below zero. For each constructed network, the connectivity density is computed as follows:
\begin{equation}
\Delta= \frac{E}{N(N-1)/2}
\label{density}
\end{equation}
where $E$ is the number of edges whose correlation coefficient has a certain relationship with a threshold and N is the number of nodes. If $\Delta=0$, all nodes are not connected; if $\Delta=1$, all nodes are connected. According to this measure, each network can be identified by a value of $\Delta$ between $0$ and $1$.

The network is constructed in the following steps:

\begin{enumerate}
\item fix the threshold $\rho_T$;
\item divide the entire observation period into daily non-overlapping windows;

\item calculate the Pearson correlation coefficient between any two nodes of the network in each daily window;

\item only select those edges whose weight satisfies the constraint on the threshold and calculate the daily connectivity density, based on the fixed ρT in each daily window.
\end{enumerate}

\begin{figure}
%%\rule{1cm}{1cm}width=\linewidth
\centering
\includegraphics[width=0.9\linewidth]{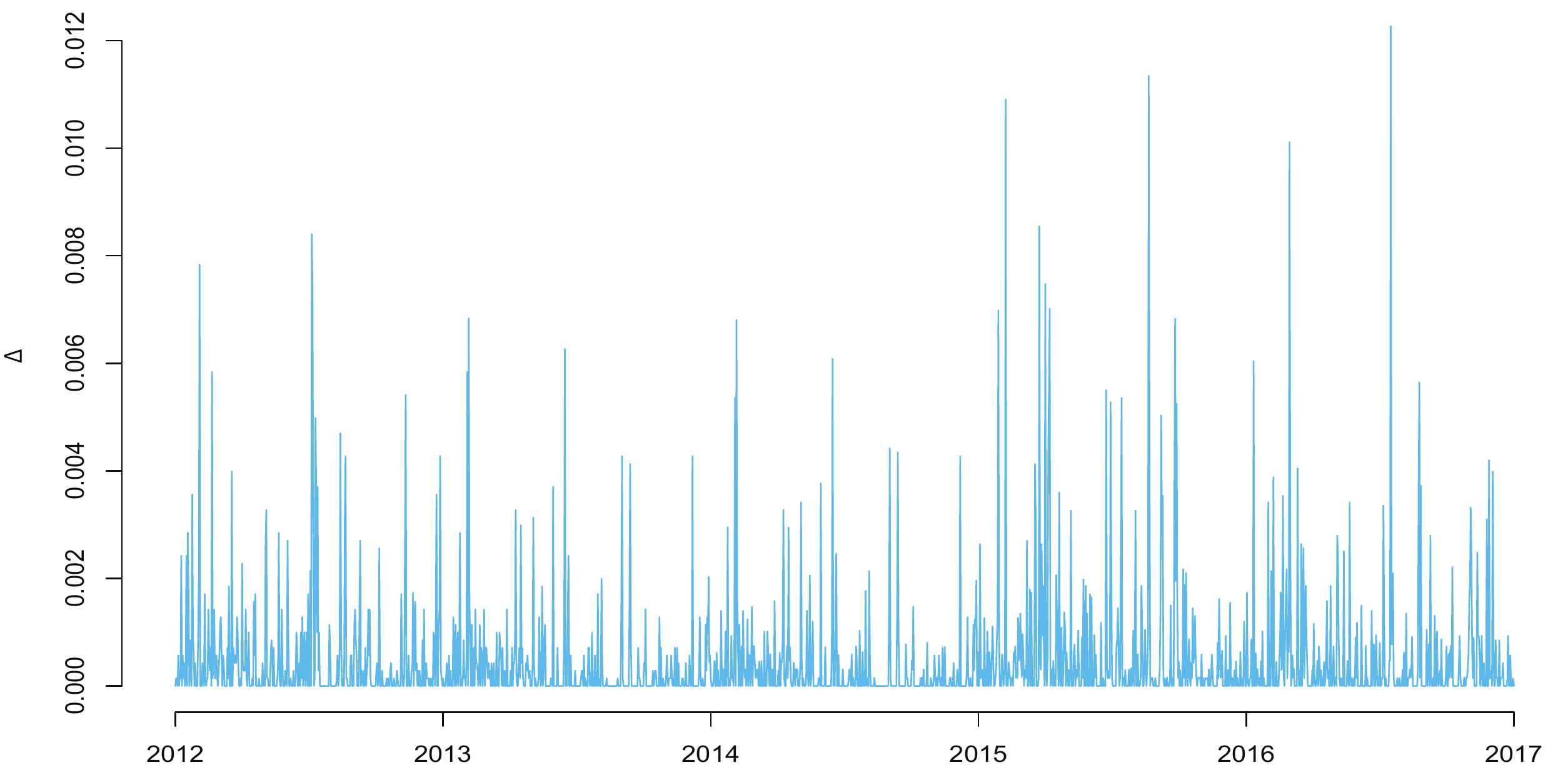}
\caption{\footnotesize Connectivity density time series for $\rho_T = -0.8$.}
\label{fig4}  
\end{figure}

\begin{figure}
%%\rule{1cm}{1cm}width=\linewidth
\centering
\includegraphics[width=0.9\linewidth]{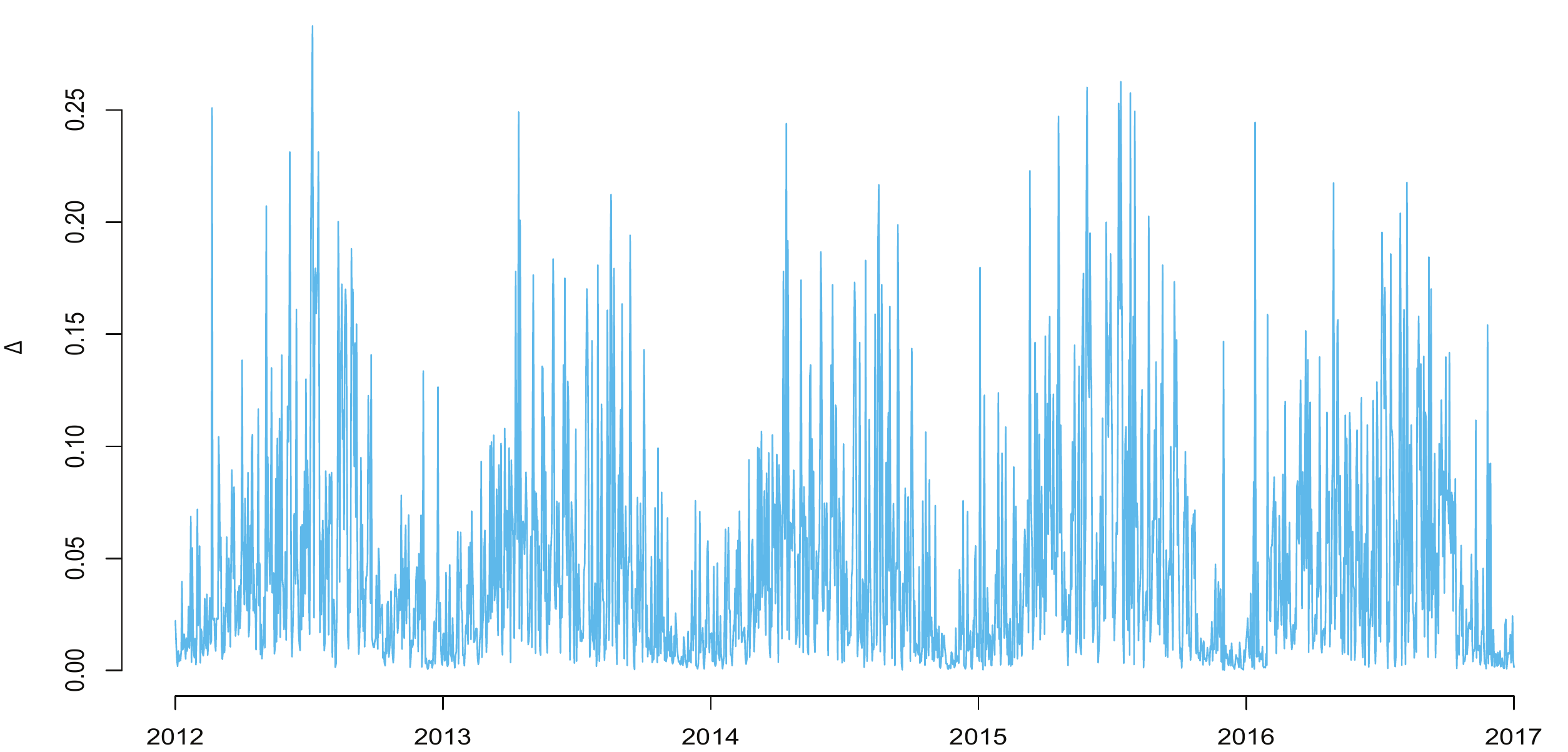}
\caption{\footnotesize Connectivity density time series for $\rho_T = 0.7$.}
\label{fig5}  
\end{figure}

Figs. \ref{fig4}   and  \ref{fig5}  show the connectivity density time series for $\rho_T= -0.8$ and $\rho_T=0.7$ respectively (The supplementary file Fig1S.pdf shows all the connectivity density time series for $\rho_T$ between $-0.9$ and $0.9$ with $0.1$ step). As it can be easily seen by visual inspection, the connectivity time series are characterized by a yearly cycle that is more visible at positive thresholds than negative ones and more intense for low absolute values of the threshold than for high absolute values.

\begin{figure}
%%\rule{1cm}{1cm}width=\linewidth
\centering
\includegraphics[width=0.8\linewidth]{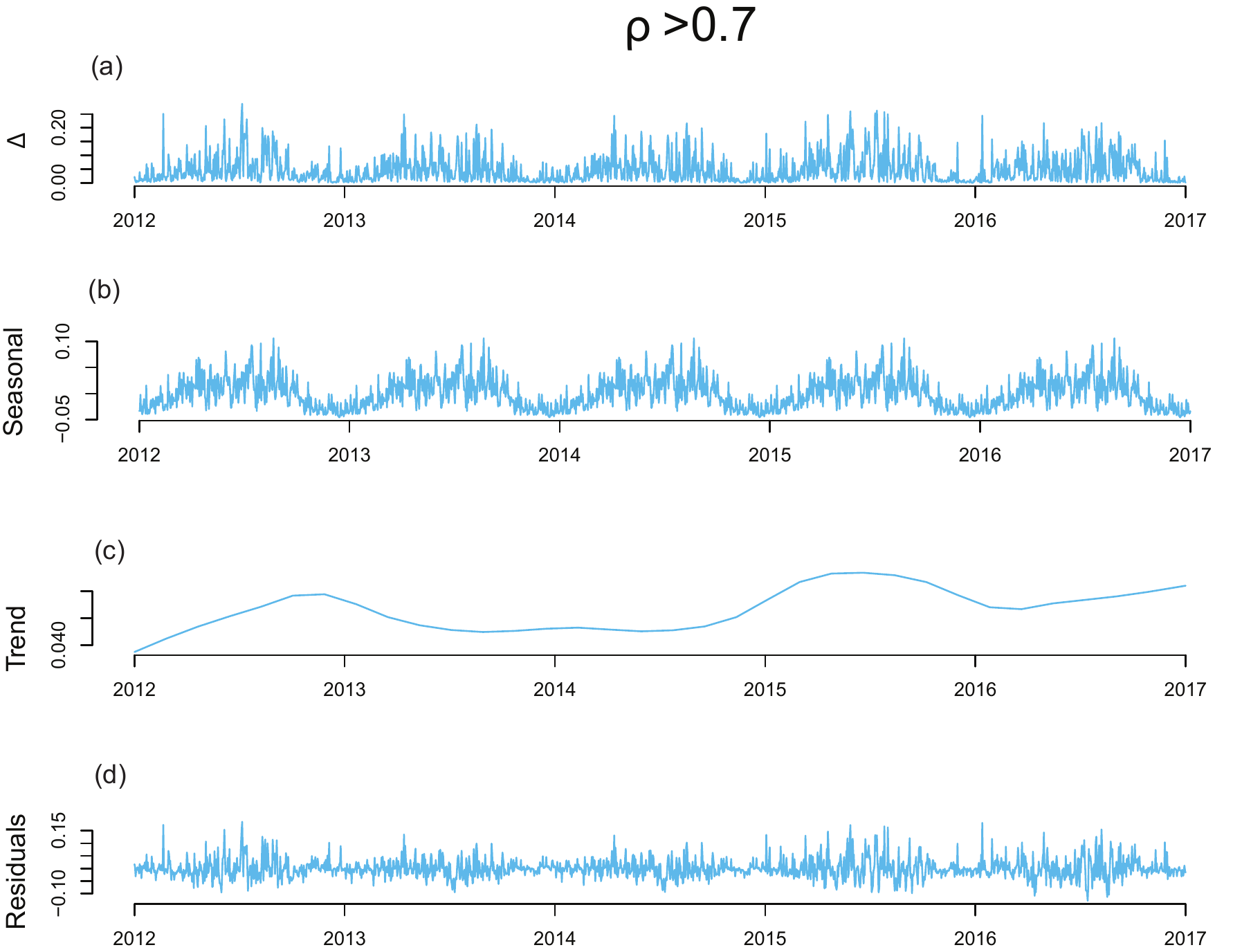}
\caption{\footnotesize STL decomposition of the connectivity time series for $\rho_T = 0.7$.}
\label{fig6}  
\end{figure}

Since the annual cycle is clearly visible in the daily connectivity time series induced by the meteo-climatic weather cycle, which can be considered as an external force, we removed it from the series before performing the MFDFA. In fact, we applied the seasonal and trend decomposition using Loess (STL) to each connectivity time series. Proposed by Cleveland \cite{Cleveland1990}, STL is a filtering procedure for decomposing a time series into trend ($T_i$), seasonal ($S_i$) and remainder (also called residuals) ($R_i$ ) components.
STL performs an additive decomposition of the global time series using the Loess smoother, which applies locally weighted polynomial regressions at each point $x_i$ of the data \cite{Theodosiou2011}.  
The overall design of the STL decomposition consists of two recursive procedures: an inner loop and an outer loop. The inner loop updates the trend and seasonal components, while the outer loop computes the robustness weights according to the remainder component, which is used in the next iteration of the inner loop. The outer loop tends to reduce the weight of outliers or extreme values in the time series. In fact, the point closest to $x_i$  has the largest weight, with the weight decreasing as the distance from $x_i$ increases. 
We applied the R stats library  (stl function) \cite{lanR} to the obtained connectivity time series (for more details on the STL decomposition procedure, refer to Cleveland et al. \cite{Cleveland1990}).
Fig.  \ref{fig6} shows as an example of the results of STL decomposition for $\rho_T=0.7$ (The supplementary file Fig2S.pdf presents the STL decomposition for all the connectivity density time series analysed in this paper).

Therefore, the connectivity time series (Fig. \ref{fig6}.a) has been divided into three components: the seasonal (Fig. \ref{fig6}.b), the trend (Fig. \ref{fig6}.c) and the remainder (Fig. \ref{fig6}.d). Among the three components, the remainder (residual) represents the daily evolution of the inner interactions of the network for a certain correlation threshold. Thus, the remainder informs how the network topology changes based on the inner relationships that emerge among the nodes.
Hereafter,  the term residual is used to intend the remainder of the connectivity time series after applying the STL. Then, the MFDFA is applied to the residuals.

\section{Multifractal detrended fluctuation analysis}

The MultiFractal Detrended Fluctuation Analysis (MFDFA) is a well-known technique, and is commonly used to detect multifractality in a time series \cite{ KANTELHARDT2002}. Let $x(i)$ for $i=1,\ldots,N$ be a possibly non-stationary time series, where $N$ indicates its length. We, first, construct the "trajectory" or the "profile" by integration after subtracting from the time series its average $x_{ave}$.
\begin{equation}
Y(i)=\sum_{k=1}^i[x(k)-x_{ave}].
\end{equation}

The profile is sub-divided into $N_s=int(N/s)$ non-overlapping windows of equal length $s$. Since the length $N$ of the series may not be an integer multiple of the window size $s$, and a short part of the profile $Y(i)$ at the end may be disregarded by the procedure, the sub-division is performed also starting from the opposite end, obtaining a total of $2N_S$ segments. A polynomial of degree $m$ fits the profile in each of the $2N_S$ windows and the variance is calculated by using the following formula:
\begin{equation}
F^2 (s,v)=\frac{1}{s} \sum_{i=1}^s\{Y[(v-1)s+i]-y_v (i)\}^2.
\end{equation}

For each segment $v$, $v=1,\ldots,N_s$  and
\begin{equation}
F^2 (s,v)=\frac{1}{s} \sum_{i=1}^s\{Y[N-(v-N_s )s+i]-y_v (i)\}^2
\end{equation}
where $v=N_(s+1),\ldots,2N_s$. Here $y_v (i)$  is the fitting polynomial in segment $v$. 

Then, averaging over all segments the $q_{th}$ order fluctuation function is computed

\begin{equation}
F_q (s)=\left\lbrace \frac{1}{2N_s} \sum_{v=1}^{2N_s}[F^2 (s,v)]^{\frac{q}{2}}\right\rbrace ^{1/q}  
\end{equation}
where, in general, the index variable $q$ can take any real value except zero. The parameter $q$ enhances the small fluctuations if negative, otherwise, the large ones if positive. $F_q (s)$ will increase with increasing $s$ and if $F_q (s)$  behave as a power-law of $s$ the series is scaling for that specific $q$. In this case

\begin{equation}
F_q (s) \propto s^{h_q}.
\end{equation}

The exponent $h_q$ is called generalized Hurst exponent due to the equivalence between $h_2$  and the Hurst exponent ($H$) \cite{Jens1988} for stationary series, leading to consider the well know detrended fluctuation analysis (DFA) \cite{PENG1995} a particular case of the MFDFA for $q=2$. For $q=0$ the value $h_0$ corresponds to the limit $h_q$ for $q \rightarrow 0$, and is obtained through the logarithmic averaging procedure:

\begin{equation}
F_0(s) \equiv exp \left\lbrace \frac{1}{4N_s} \sum_{v=1}^{2N_s} ln[F^2 (s,v)]\right\rbrace  \propto s^{h_0}.
\end{equation}

In general, the exponent $h_q$ will depend on $q$, and it monotonically decreases with the increase of $q$, the series is multifractal. If $h_q$ does not depend on $q$ the series is mono-fractal.
Multifractal series can be also studied by means of the singularity spectrum, obtained applying the Legendre transform \cite{Parisi1983}. From the relationship

\begin{equation}
\tau(q)=qh_q-1
\end{equation}
and 
\begin{equation}
\alpha=d\tau/dq
\end{equation}
we obtain
\begin{equation}
f(\alpha)=q\alpha-\tau(q)
\end{equation}
where $\alpha$ is the H\"older exponent and $f(\alpha)$ indicates the dimension of the subset of the series \cite{Ashkenazy2003} that is characterized by $\alpha$. The multifractal spectrum indicates how much dominant are the various fractal exponents present in the series. The width of the singularity spectrum, as well as the range of the generalized Hurst exponent $(max(h_q)-min (h_q))$, are often used to quantitatively measure the degree of multifractality of the series. Thus, the wider the spectrum the more multifractal the series is. The R codes used for this study are available on GitHub as an R library \cite{MFDFA2017}.

\section{Results}
The MFDFA was applied to all the residuals for correlation thresholds $\rho_T$ between $-0.9$ to $0.9$. Since each time series has 1825 daily samples, we considered   time scales up to 180 days.

\begin{figure}
%%\rule{1cm}{1cm}width=\linewidth
\centering
\includegraphics[width=0.9\linewidth]{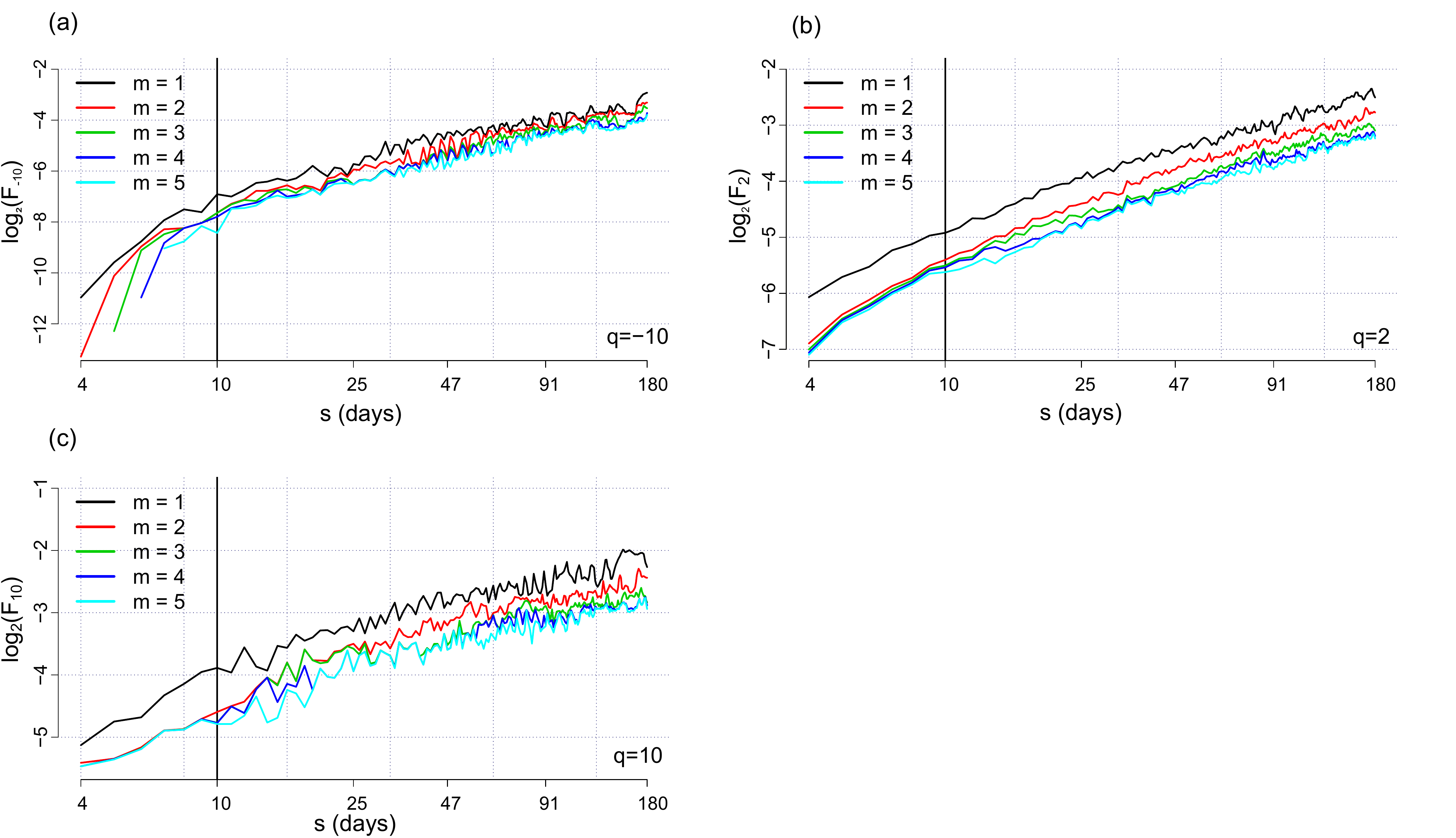}
\caption{\footnotesize MFDFA results for  $m=1,\ldots,5$ and for $\rho_T = 0.7$.}
\label{fig7}  
\end{figure}

In order to remove all the non-stationarities in the residuals, it is important to determine the degree of the polynomial in the MFDFA. As several studies \cite{KANTELHARDT2002, BISWAS2012} have reported, the optimal degree (m) of the detrending polynomial can be guessed by first comparing the fluctuation functions for different values of m. And second, by selecting the m whose fluctuation function overlaps with the fluctuation function  of superior degree or that has very close value of the slope of the linear fitting function .  
Fig. \ref{fig7} shows fluctuation functions for $F_{-10}$ (Fig. \ref{fig7}.a), $F_2$ (Fig. \ref{fig7}.b), $F_{10}$ (Fig. \ref{fig7}.c) for $\rho_T=0.7$ for scales from $4$ to $180$ for $m=1,\ldots,5$.
From visual inspection, it is very clear that the fluctuation functions for $m=3$ almost overlap those for $m=4$, suggesting that a third-degree polynomial would perform the optimal detrending. Therefore, we will use $m=3$; this value is also in agreement with  Oświęcimka et al. \cite{OSWI2013}, who found that $m=3$ has to be preferred in order to avoid the bias in the multifractal results. Moreover, especially for negative $q$ (Fig. \ref{fig7}.a), the behaviour of the fluctuation function, plotted in log-log scales, is not linear at small scales of up to 10 days; therefore, the MFDFA  is  performed in the scale range between 10 and 180 days.

The analysis of $F_2$ allows us to obtain information about the persistence of the series. The persistence of a time series has been standardly recognised using an analysis of the power spectrum. A white noise, that is typical of purely random temporal fluctuation, is characterised by a power spectrum approximately constant for any frequency band,  which means that any sample is totally independent of the others (the series is deprived of any memory feature). If the power spectrum behaves as a power-law function of the frequency, this suggests that long-range correlation exists in the series. In this case, the value of the spectral exponent $\alpha$ furnishes relevant information about the temporal fluctuations of the series  \cite{MANDELBROT1983, Schertzer1987, Marshak1994, Calif2014}: 
If $\alpha < 1$, the series is stationary; if $-1 < \alpha < 0$, the series is anti-persistent (the series is characterized by an apparent alternation of  increments and decrements); if $\alpha=0$, the series is uncorrelated; if $0< \alpha < 1$ the series is persistent (increments (decrements) of the series tend to follow increments (decrements)  of the series). If $\alpha > 1$, the series is non-stationary; in particular if $1 < \alpha < 3$, the series is non-stationary with stationary increments. 

In case of stationary series, the spectral exponent relates to the Hurst exponent $H$ (a well-known parameter used to describe the persistence of a series) \cite{parzen1986} by $\alpha=2H-1$. Thus, the persistence/anti-persistence of the series could be deduced from H (larger or smaller than $0.5$ that represents the value of $H$ for pure randomness). Although the power spectrum represented the standard method for evaluating the persistence of the series. The possible non-stationarities of unknown origin that often affect observational data renders the use of methods that are more robust than the power spectrum necessary. 
Due to the robustness of the MFDFA to detect scaling in non-stationary series, $F_2$ (whose $h_2$ exponent coincides with the Hurst exponent $H$ for stationary series) can be employed to investigate the persistence in our residuals. 

\begin{figure}
%%\rule{1cm}{1cm}width=\linewidth
\centering
\includegraphics[width=0.9\linewidth]{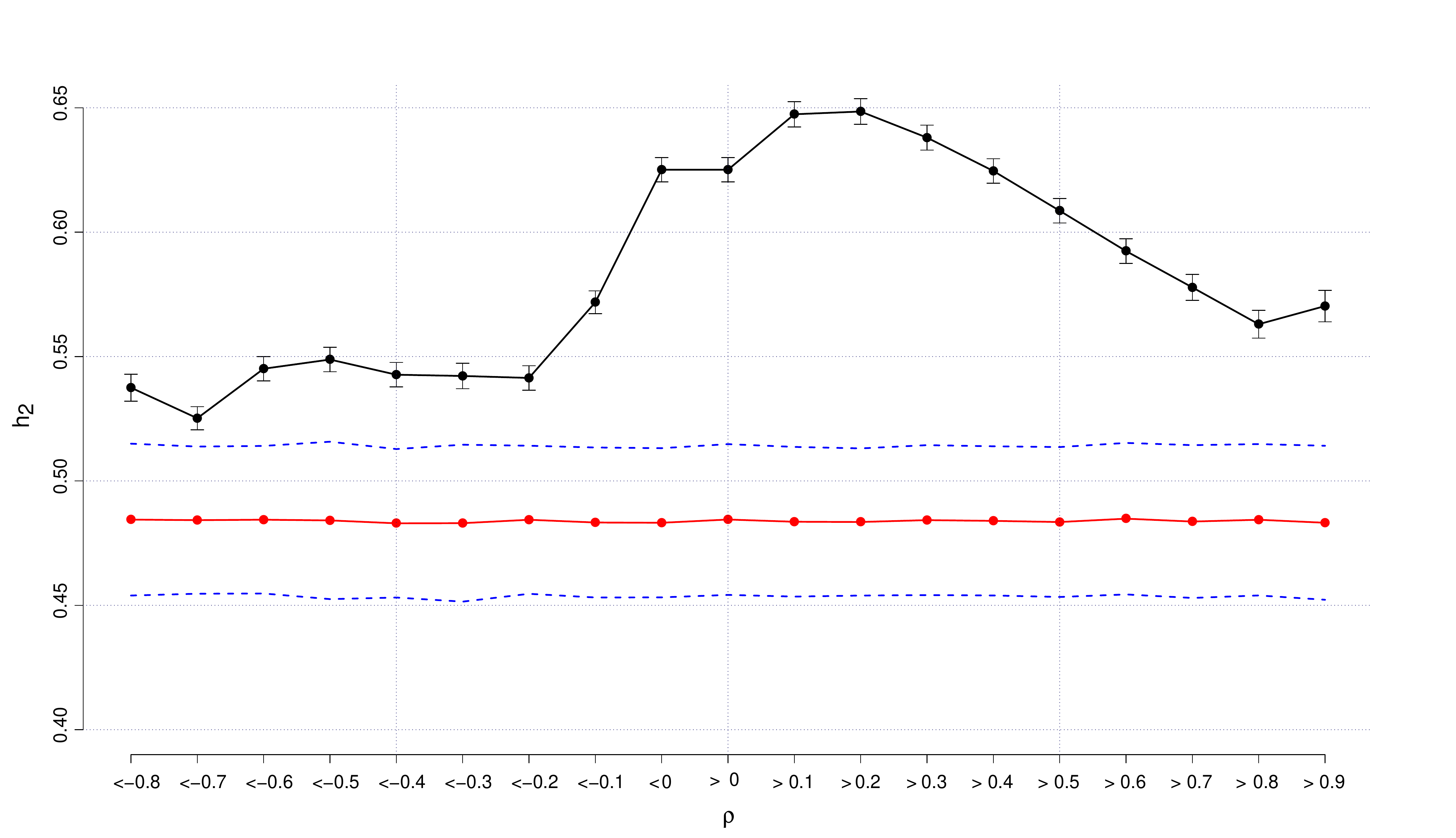}
\caption{\footnotesize $h_2$ exponents for all the analysed residuals (black); average (red) and standard deviation band (blue) over 20,000 $h_2$ exponents calculated for the shuffles.}
\label{fig8}  
\end{figure}

Fig. \ref{fig8}  shows the variation of the $h_2$ exponent with the threshold $\rho_T$. The Hurst exponent $H$ varies between $0.525 \pm 0.004$ and $0.649 \pm 0.005$, indicating that the residuals are persistent for any threshold. In order to check the results against the randomness, for each series we generated $20,000$ shuffles and for each shuffle we calculated the $h_2$ exponent with $m=3$ in the same time scale range as for the original series. The shuffling saves the distribution of the series but destroys all the correlation structures. Fig. \ref{fig8} also shows the average over the $20,000$ $h_2$ exponents of the shuffles within $1$ standard deviation band for each threshold. As is clearly visible, the residuals are significantly persistent for any $\rho_T$, being maximally persistent for $\rho_T =0.1$.

\begin{figure}
%%\rule{1cm}{1cm}width=\linewidth
\centering
\includegraphics[width=\linewidth]{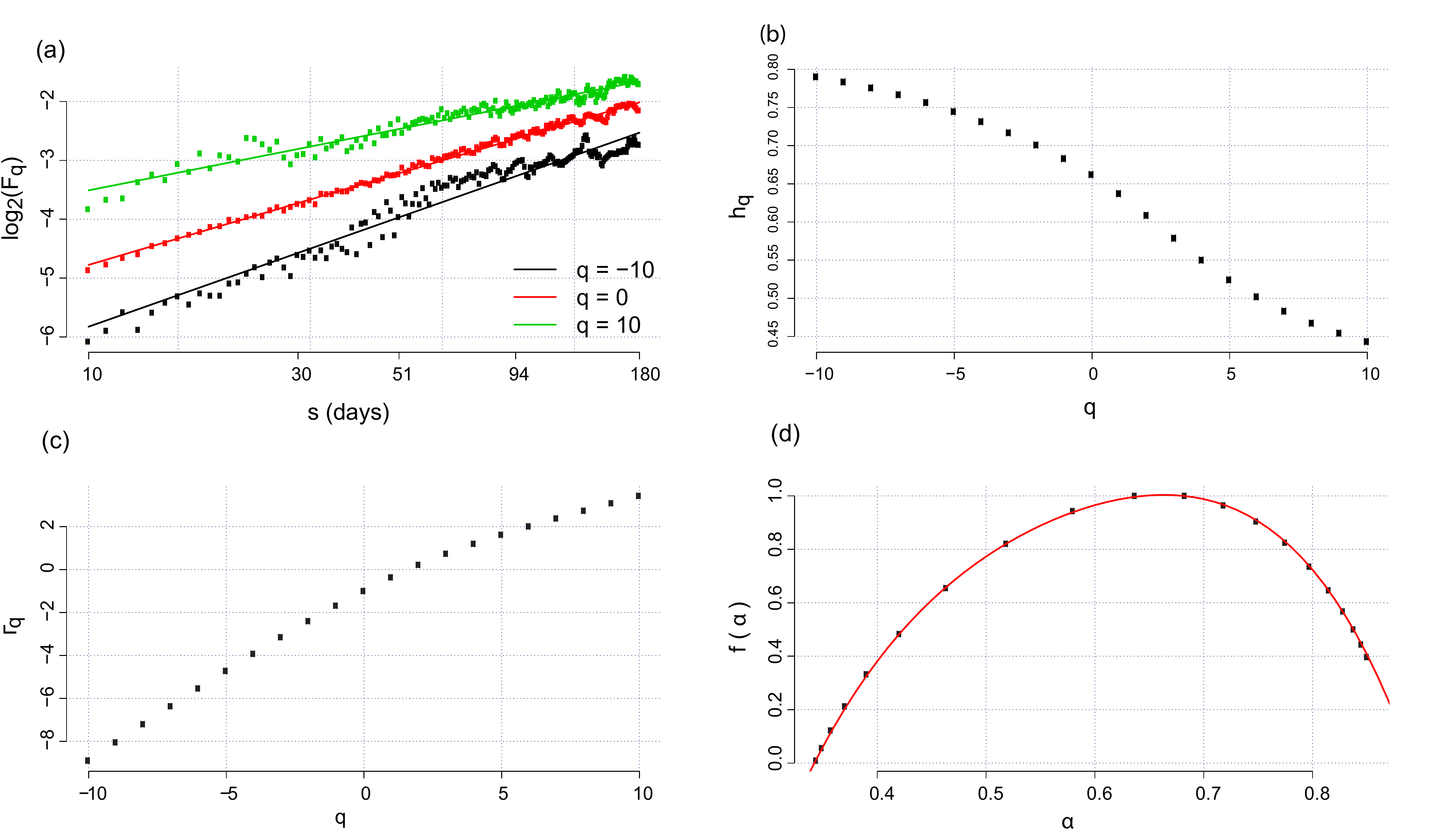}
\caption{\footnotesize MFDFA results for $\rho_T =0.5$; (a) fluctuation functions; (b) generalized Hurst exponents; (c) $\tau$-function; (d) multifractal spectrum.}
\label{fig9}  
\end{figure}

We applied the MFDFA to the residuals for $q$ ranging between $-10$ and $10$ with $1$-step, for $m=3$ and time scales between $10$ days and $180$ days. Fig. \ref{fig9} demonstrates as an example for $\rho_T =0.5$ the fluctuation functions for $q=-10$, $q=0$ and $q=10$ (Fig.  \ref{fig9}.a), the generalized Hurst exponents (Fig. \ref{fig9}.b), the function $\tau(q)$ (Fig. \ref{fig9}.c) and the multifractal spectrum (Fig. \ref{fig9}.d) (the supplementary Fig4s.pdf shows the results for all the thresholds $\rho_T$). For $\rho_T =-0.9$ the fluctuation functions appear quite unstable for negative $q$ and $a$ multifractal behaviour cannot be recognized; thus all the MFDFA results are presented for thresholds larger than $-0.9$. 

\begin{figure}
%%\rule{1cm}{1cm}width=\linewidth
\centering
\includegraphics[scale=0.5]{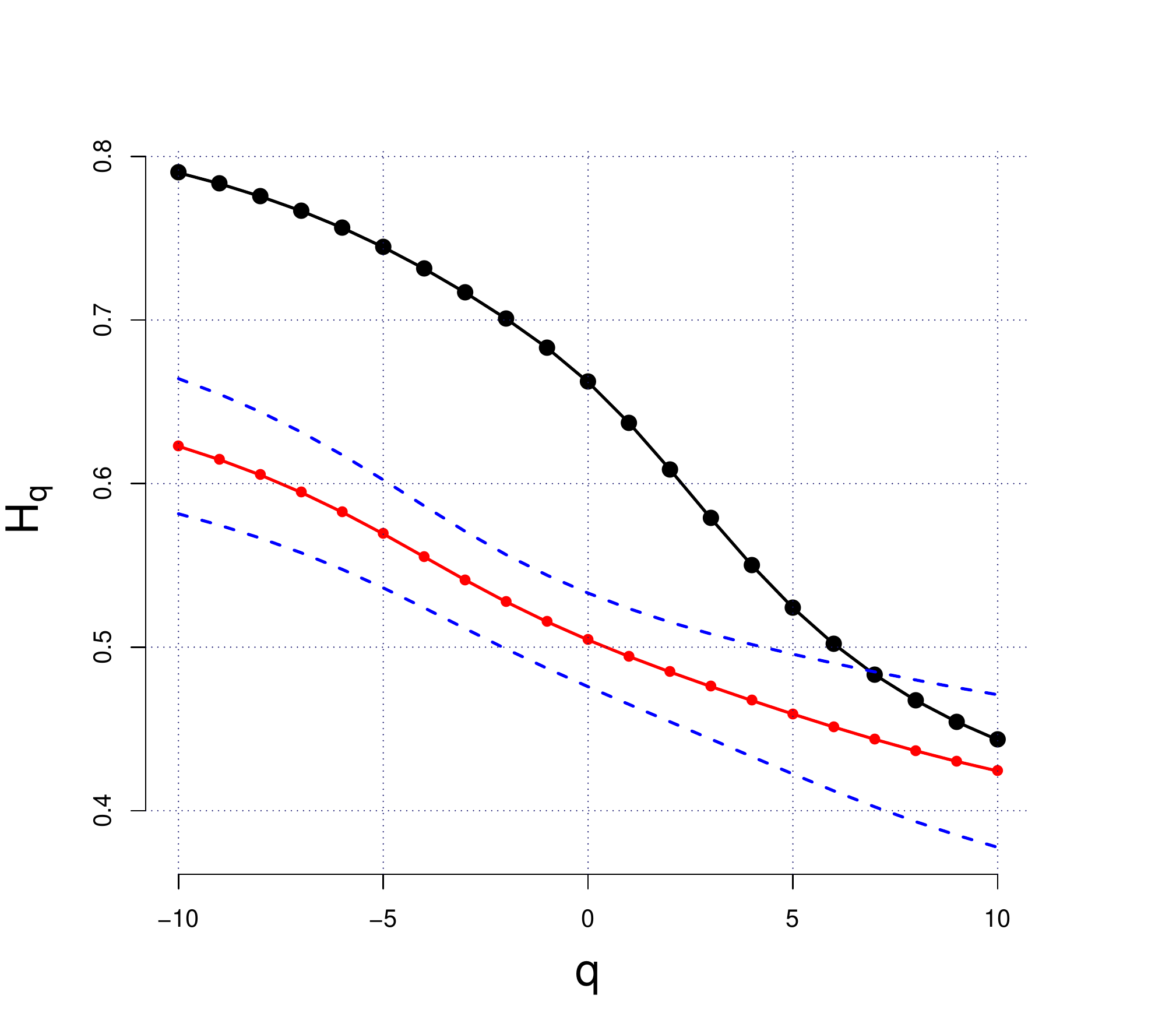}
\caption{\footnotesize Generalised Hurst exponent for analysed residuals (black) and the average (red) over $20,000$ shuffled residuals time series. (For $\rho_T=0.5$).}
\label{fig10}  
\end{figure}

\begin{figure}
%%\rule{1cm}{1cm}width=\linewidth
\centering
\includegraphics[width=0.9\linewidth]{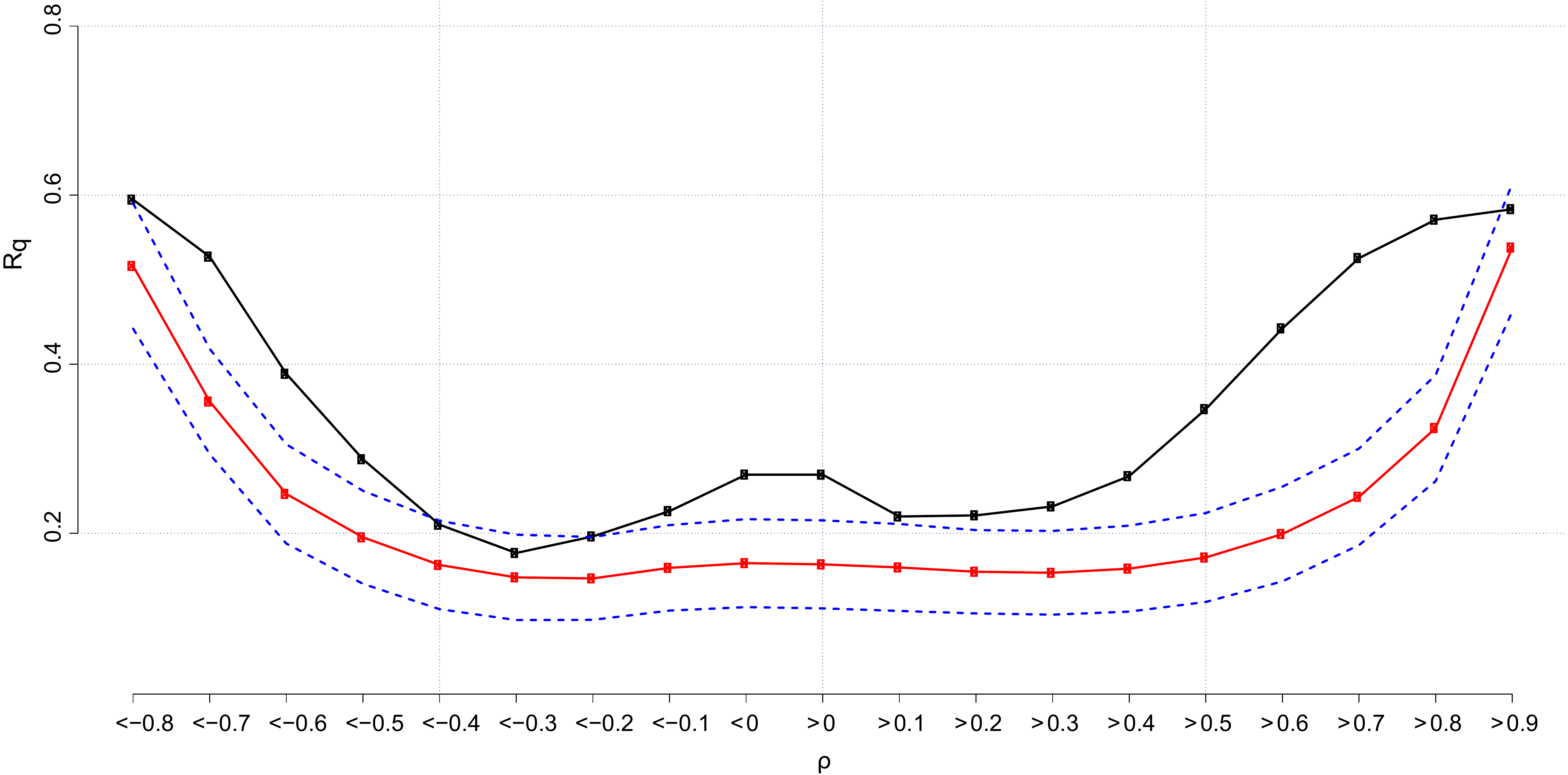}
\caption{\footnotesize Range of the Hurst exponent for analysed residuals (black); average (red) and standard deviation band (blue) over $20,000$ shuffled residuals time series.}
\label{fig11}  
\end{figure}

\begin{figure}
%%\rule{1cm}{1cm}width=\linewidth
\centering
\includegraphics[width=0.9\linewidth]{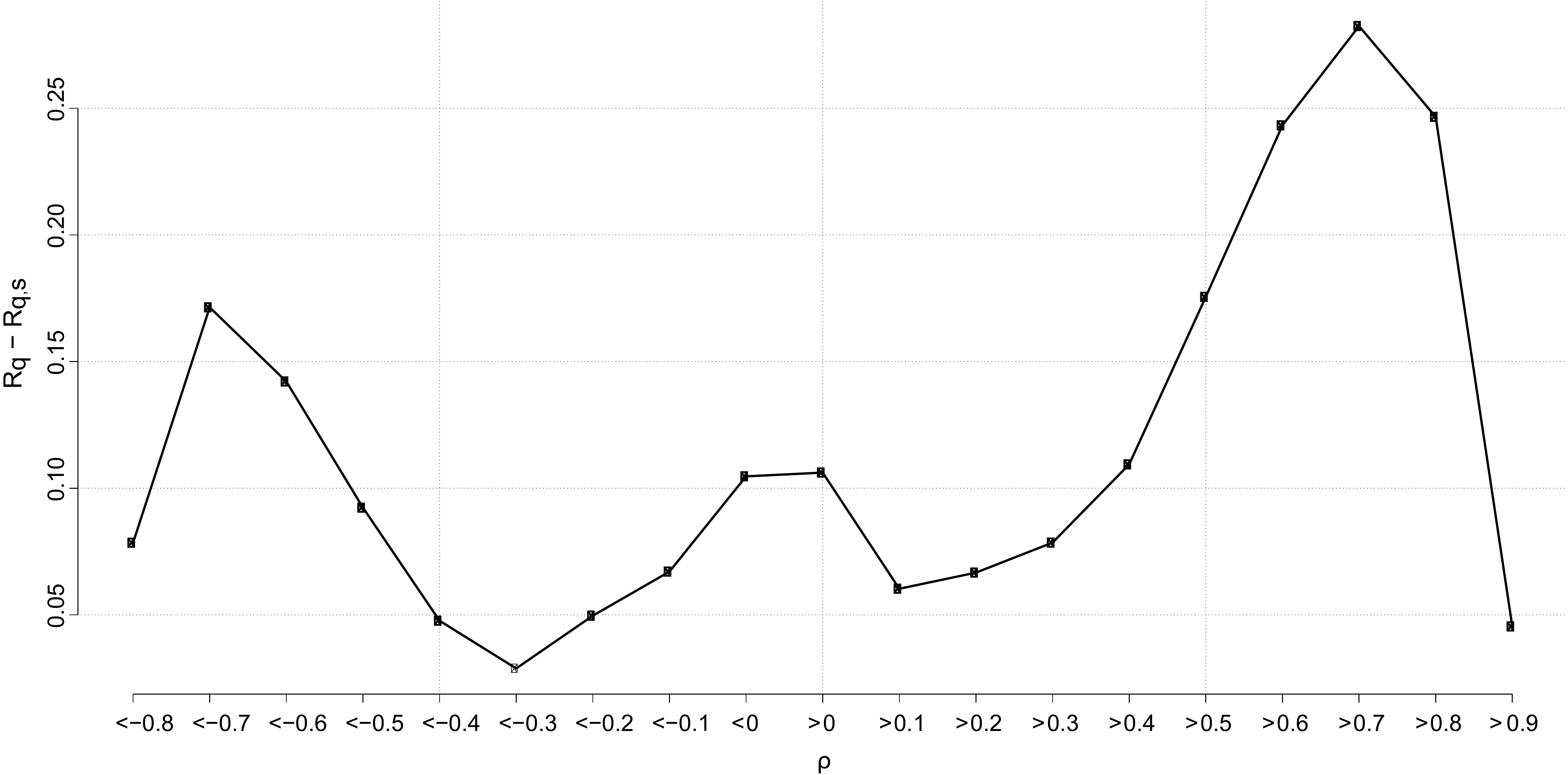}
\caption{\footnotesize Difference between $R_q$ and $R_{q,s}$.}
\label{fig12}  
\end{figure}

The residuals are clearly multifractal, since the generalized Hurst exponents are not approximately constant for any $q$, but decrease with the increase of $q$, and the multifractal spectrum has its typical single-humped shape that characterizes a series with multifractal behaviour. In order to see whether found multifractality depends on the long-range correlations or on the probability density function of the series, we generated $20,000$ shuffles of the residuals and applied the MFDFA to each shuffle. Fig. \ref{fig10}  shows the comparison between the generalized Hurst exponents of the original residual time series for $\rho_T =0.5$ and the mean ($\pm$ $1$ standard deviation) generalized Hurst exponent calculated over $20,000$ shuffles. The generalized Hurst exponents are not approximately constant for any $q$, but are characterized by a slight decrease with an increase of $q$, indicating that the observed multifractality also depends on the distribution of the residuals. This effect is present in all the residuals (see the supplementary file Fig5S.pdf); however, the dependence of the multifractality upon the distribution is not the same for all the thresholds.
Fig.  \ref{fig11} shows the range  $R_q$  of the generalized Hurst exponent (difference between the maximum and the minimum $h_q$ ) . Fig. \ref{fig12}  shows the difference between $R_q$ of the original residuals and $<R_{q,s}>$, which is the mean range of generalized Hurst exponents calculated over $20,000$ shuffles. 
For each threshold, the maximum difference is for $\rho_T =0.7$, suggesting that for this value of the correlation threshold the distribution would have the least effect in driving the multifractality in the residual.

\begin{figure}
%%\rule{1cm}{1cm}width=\linewidth
\centering
\includegraphics[width=0.9\linewidth]{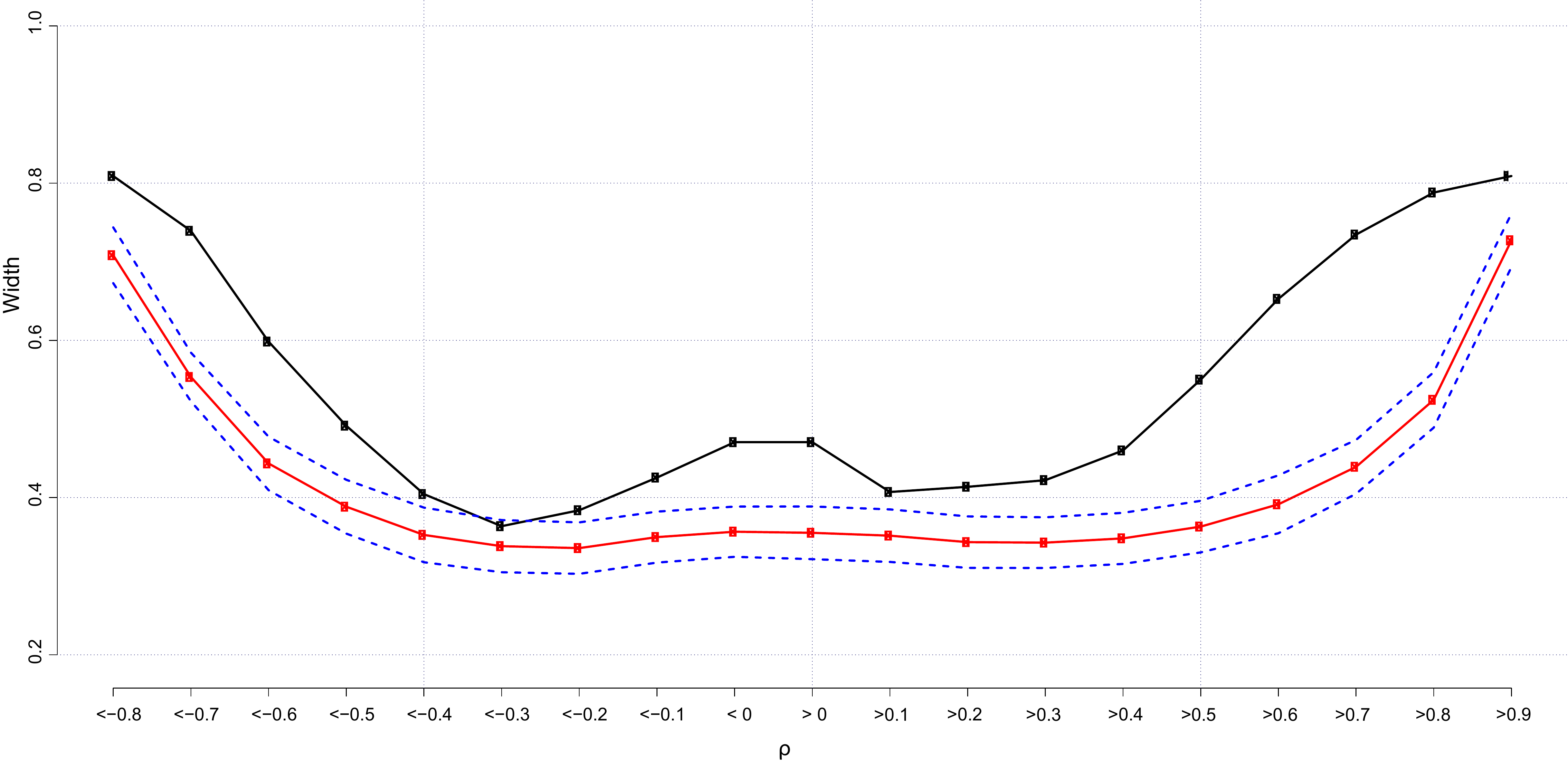}
\caption{\footnotesize Width of the multifractal spectrum curve(black); average (red) and standard deviation band (blue) over $20,000$ shuffled residuals time series.}
\label{fig13}  
\end{figure}

To further quantify the degree of multifractality (also indicated by $R_q$) we fitted the multifractal spectrum with a $4^{th}$ order polynomial and calculated the width of the spectrum by the distance between the two zero-crossings of the fitting function. Fig. \ref{fig13} shows the width for each threshold,
revealing that the multifractality degree of the residual increases with the increase of the absolute value of the threshold.

\section{Conclusion}
\begin{enumerate}
\item The connectivity density time series are characterized by an evident annual cycle that becomes weaker with the increase of the absolute value of the correlation threshold.

\item The application of the STL algorithm allows the residuals to be extracted, and reveals the inner interactions among the nodes of the wind network. The regularity inherited in the seasonal component can describe the annual weather-induced component of the connectivity density, while the residual can likely describe the local meteo-climatic conditions (which depend upon the geomorphology, sun exposure, altitude, etc) that characterize the site of each measuring station.

\item The analysis of the Hurst exponent, derived from the slope of the fluctuation function corresponding to $q = 2$, suggests that the residuals are persistently correlated for any threshold. Further, there are long-range correlation structures in the topology evolution  of networks. In addition, the Hurst exponent is lower for negative thresholds than for positive ones. Such network asymmetry in the correlative domain (since the strength of the long-range correlation is weaker for negative thresholds than for positive) suggests that the in-phase fluctuations have stronger memory phenomena than the anti-phase fluctuations;

\item Since the residuals mainly describe the local meteo-climatic-induced fluctuations of the network topology, it will be challenging in future studies to link the in-phase and anti-phase fluctuations (corresponding to the positive and negative correlation thresholds respectively) to the local site features (like geomorphology, altitude, sun exposure, etc.) in order to find which feature would be more dominant in the in-phase or anti-phase fluctuations;

\item The multifractality degree is higher for larger absolute values of the correlation threshold. The multifractality degree is a measure of the heterogeneity of the residuals. Therefore, a higher multifractality degree (larger heterogeneity) means that the residuals intermittency is more intense. There seems to be no difference between positive and negative thresholds, since the variation of the width of the multifractal spectrum with the threshold is approximately symmetric. The larger multifractality at higher absolute values of thresholds could probably reflect the higher spatial sparseness of the linked nodes at these thresholds. At low absolute values of the threshold, more nodes are interconnected by the correlation coefficient. Furthermore, the spatial configuration of the network appears denser and more homogeneously structured. At larger absolute values of the threshold, less nodes are interconnected;

\item In conclusion, the results presented in this paper enable us to envisage a novel perspective in the context of studies devoted to the analysis of wind speed time series. Focusing on the cooperative behaviour of wind speed monitoring systems as a correlated network could contribute to a better understanding of the mechanisms underlying the variability of wind speed.

\item The present work studies a wind monitoring system and proposes an approach based on the correlation networks. Further studies could be in introducing the geo-spatial coordinates, and applying the similar approach including the concept of spatial networks that would be very useful in the design of the monitoring system.
\end{enumerate}

\section{Acknowledgements}
The authors thank MeteoSwiss for the accessibility to the data via the IDAWEB server. They also are grateful to the anonymous reviewers for their constructive comments that contributed to improving the paper.

This research was partly supported by the Swiss Government Excellence Scholarships. LT thanks the support of Herbette Foundation.

\section*{Supplementary Material}

In order to better explain the present research and since the paper studies several time series, we added Supplementary On line Material.

\subsection*{Fig1S.pdf}
Contains the connectivity density times series for different thresholds $\rho_T$.
\subsection*{Fig2S.pdf}
Contains the STL decomposition for all connectivity density time series for different $\rho_T$.
\subsection*{Fig3S.pdf}
MFDFA results for different thresholds $\rho_T$.

\subsection*{Fig4S.pdf}
Generalised Hurst exponent for analysed residuals with the average of over 20,000 shuffled residuals time series (for all thresholds).
%\includepdf[fitpaper=true,pages=-]{supp/Fig1S.pdf}

%\nocite{*}

\bibliography{aipsamp}% Produces the bibliography via BibTeX.

%\cleardoublepage
%fig 11

%\cleardoublepage

%\includepdf[fitpaper=true,pages=-]{supp/Fig2S.pdf}

%\includepdf[fitpaper=true,pages=- ]{supp/Fig3S.pdf}

%\includepdf[fitpaper=true,pages=- ]{supp/Fig4S.pdf}

\end{document}